# Theoretical modelling of arch-shaped carbon nanotube resonators exhibiting Euler-Bernoulli snap-through bi-stability

S. Rechnitz, S. Shlafman, T. Tabachnik, M. Shlafman, and Y. E. Yaish[*]

*Andrew and Erna Viterbi Faculty of Electrical Engineering, Technion, Haifa, Israel.*

**Abstract**

In this work, we present a detailed static and dynamic analysis of a recently reported electrically actuated buckled carbon nanotube (CNT) resonator, based on the Euler-Bernoulli beam theory. The system behavior is analyzed using the Galerkin reduced order model. We show that a simple single-modal analysis can already predict snap-through bi-stability in a buckled CNT resonator. However, we prove that the experimental data, in which the snap-through buckling occurs at a finite frequency, cannot be explained without taking into account out-of-plane motion. The buckled CNTs are the first type of buckled beams to exhibit out-of-plane static motion, resulting in a unique three-dimensional snap-through transition, never before predicted. In addition, we show the criteria under which these devices can also exhibit latching phenomena, meaning that they can maintain their buckle configuration when no force is applied, making these devices appealing for mechanical memory applications.

# 1. Introduction

Almost everywhere in our daily life we encounter micro-electro-mechanical systems (MEMS) technology by a large number of devices with a wide range of applications[1-3]. These devices are frequently found in electrical systems as filters, phase shifters, resonators, and radio-frequency (RF) switches[4,5]. In biological and chemical systems, they serve as ultra-sensitive gas and mass detectors, reaching the scale of small molecules or single cells[6]. In mechanical systems, they play an important role in accelerometers in air bag mechanism in vehicles, pressure sensors, actuators, gyroscopes, and much more[7-9].

Common MEMS designs are based on electrostatically actuated initially curved micro-beams[10-12]. These devices possess diverse static behavior (snap-through (ST), pull-in, latching) as well as intriguing linear and nonlinear dynamics (mode-coupling, veering, harmonics, internal resonance)[13-17].

These phenomena are characterized by large static and dynamic behavior of the device due to small variations in its geometric or physical parameters, and have been widely studied in the context of both potential applications as well as fundamental understanding of mode coupling and nonlinear phenomena.

As fabrications capabilities and techniques are improved, smaller mechanical objects were fabricated, entering the regime of nano-electro-mechanical systems (NEMS)[18]. These miniaturized devices are able to achieve even better performance with respect to their MEMS counterparts[19]. Indeed, fabrication of bistable buckled nano beams has also been demonstrated[20].

Recently, we reported the fabrication of a bistable NEMS device based on a buckled suspended carbon nanotube (CNT), exhibiting snap-through buckling bi-stability[21]. We showed that in addition to their small dimensions and record frequency electrostatic tunability, these devices exhibit a new type of mode coupling which results from a three-dimensional static motion, which was not possible in traditional micro and nano beams, mainly due to the large differences in the resonance vibrational modes of the out-of-plane and in-plane motion.

In this work, we present a full theoretical study of the new mechanics enabled by these buckled nanotubes, based on coupled Euler-Bernoulli (EB) beam equations for the in-plane as well as the out-of-plane static and dynamic motions. We use the Galerkin reduced-order model (ROM) to solve these equations, validated by a finite element method (FEM) approximation. We emphasize the two main differences of the buckled CNT devices compared to traditional micro-beams: (1) The force is inversely proportional to the displacement (resulting from the capacitance model between a wire and a plane), and (2) the three-dimenssional static motion. We show that after developing the full 3D model, we can predict a rotational continuous motion downward, a 3D EB ST buckling transition, or latching, depending on the initial configuration of the device. This model was used for the theoretical fitting in Ref. 21 and produces an excellent agreement with the experimental data.

## 2. Problem formulation - 2D model

We begin by modelling the system as a standard doubly clamped beam, as illustrated in Fig. 1.

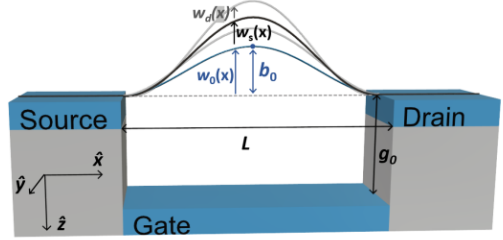

**Figure 1 Device schematic.** *Schematic illustration of an initially upward buckled CNT device. w(x) is the CNT z displacement.*

Neglecting damping and limiting the CNT motion to the xz-plane, it can be described by the Euler-Bernoulli beam equation:

$$EI\frac{\partial^4 \hat{w}}{\partial \hat{x}^4} + \rho A \frac{\partial^2 \hat{w}}{\partial \hat{t}^2} - T\frac{\partial^2 \hat{w}}{\partial \hat{x}^2} = \hat{\kappa} \quad (1)$$

where $\hat{x}$ is the axis along the beam, $\hat{t}$ is time, $\hat{w}(\hat{x}, \hat{t})$ is the deflection along the beam in the z-axis, $E$ is the Young's modulus, $I$ is the moment of inertia, $\rho$ is the mass density, and $\hat{\kappa}$ is the electrostatic force exerted by the local-gate along the z axis per unit length. $T$ is the tension given by[22]

$$T = T_0 - \frac{EA}{2L}\int_0^L \left(\frac{\partial \hat{w}}{d\hat{x}}\right)^2 d\hat{x} \quad (2)$$

in which $L$ is the CNT length, $A$ is the CNT cross section area, $T_0$ is the initial axial tension when no external force is applied, and the second term presents the contribution of the

CNT's motion to the induced built-in tension. All simulations in this chapter are performed for a device with the physical parameters detailed in Table 1, unless specified otherwise.

| Symbol | Physical interpretation | Value for simulation |
|---|---|---|
| **L** | CNT length | 1 μm |
| **R** | CNT cross-section radius | 1 nm |
| **g$_0$** | CNT height above the local gate | 150 nm |
| **E** | CNT Young's modulus | 3.5 TPascal[26] |
| **ρ** | CNT mass density | 1300 kg/m³ [27] |

**Table 1** *Physical parameters of a typical device simulated in this work.*

*2.1 Electrostatic force*

The force acting on the CNT per unit length, exerted by the local gate (LG), is given by:

$$\hat{\kappa} = \frac{1}{2}\frac{\partial C_g}{\partial z}\left(V_0 + V_{gDC} + V_{gAC}\right)^2 \quad (3)$$

where $V_{gDC}$ and $V_{gAC}$ are the DC voltage and AC actuation amplitude applied to the local gate, respectively. Since the harmonic excitation is small ($V_{gAC} \ll V_{gDC}$), we neglect the quadratic term, such that:

$$\left(V_0 + V_{gDC} + V_{gAC}\right)^2 \approx \left(V_0 + V_{gDC}\right)^2 + 2\left(V_0 + V_{gDC}\right)V_{gAC} \quad (4)$$

In general, we assume that there may exist a small electrostatic force between the LG and the CNT even at $V_{gDC}=0$[23,24], due to the work function difference between the CNT and metal electrodes, resulting in transfer of charges to or from the CNT upon contact. This effect is represented by the $V_0$ term, and its significance will be clarified in the following section.

The capacitance between the CNT and the local gate (per unit length) is taken as the capacitance of a wire parallel to a plane, and assuming $\frac{(g_0+w)^2}{r^2} \gg 1$, it can be approximated as:

$$C_g(z) = \frac{2\pi\varepsilon_0}{\ln\left(\frac{2(g_0+\hat{w})}{r}\right)} \qquad (5)$$

where $\varepsilon_0$ is the vacuum permittivity, r is the CNT radius, and $g_0$ is the LG-CNT distance when the tube is straight and w=0. The derivative is therefore given by:

$$\frac{\partial C_g}{\partial z} = \frac{\pi\varepsilon_0}{(g_0+\hat{w})\left(\ln\left(\frac{2(g_0+\hat{w})}{r}\right)\right)^2} \qquad (6)$$

In typical devices, $\left|\frac{\hat{w}}{g_0}\right| < \frac{1}{3}$, and since the tube radius is approximately 1nm, $\ln\left(\frac{2(g_0+\hat{w})}{r}\right) = \ln\left(\frac{2g_0}{r}\cdot\left(1+\frac{\hat{w}}{g_0}\right)\right) \approx \ln\left(\frac{2g_0}{r}\right) + \frac{\hat{w}}{g_0} \approx \ln\left(\frac{2g_0}{r}\right)$. Therefore, we choose to neglect the CNT displacement inside the log argument. Justification for this approximation is presented in section 3.3.

We shall emphasize a significant difference between this analysis and previous studies on buckled beams[25,26,30,31]. In those studies the capacitance derivative between two plates is inversely proportional to the displacement squared ($F \propto 1/z^2$)[26], whereas in our case, it is inversely dependent on the displacement ($F \propto 1/z$).

To conclude, our initial 2D model is based on the following EB beam equation:

$$EI\frac{\partial^4 \hat{w}}{\partial \hat{x}^4} + \rho A \frac{\partial^2 \hat{w}}{\partial \hat{t}^2} - T\frac{\partial^2 \hat{w}}{\partial \hat{x}^2} = \frac{1}{2}\frac{\pi\varepsilon_0}{(g_0 - \hat{w})\left(\ln\left(\frac{2g_0}{r}\right)\right)^2}\left((V_0 + V_{gDC})^2 + 2(V_0 + V_{gDC})V_{gAC}\right) \quad (7)$$

*2.2 Nondimensional beam equation*

It is customary to work in dimensionless units. All non-dimensional parameters used in the formulation are detailed in Table 2.

| Mathematical definition | Physical interpretation |
|---|---|
| $x = \hat{x}/L$ | Normalized coordinate along the CNT |
| $w = \hat{w}/g_0$ | Normalized movement in the xz-plane relative to the CNT elevation above the local gate, $g_0$ |
| $P = T_0 L^2 / EI$ | Initial axial strain |
| $\alpha = A g_0^2 / 2I$ | Axial force parameter |
| $\kappa_0 = \pi\varepsilon_0 / g_0 \ln^2(2g_0/r) \cdot L^4 / EIg_0 \cdot V_0^2$ | Initial electrostatic force at Vg=0 |
| $\kappa_s = \pi\varepsilon_0 / g_0 \ln^2(2g_0/r) \cdot L^4 / EIg_0 \cdot (V_0 V_{gDC} + V_{gDC}^2)$ | External electrostatic static force (DC) |
| $\kappa_d = \pi\varepsilon_0 / g_0 \ln^2(2g_0/r) \cdot L^4 / EIg_0 \cdot 2(V_0 + V_{gDC})V_{gAC}$ | External electrostatic harmonic actuation |
| $t = \hat{t}\sqrt{EI/\rho A L^4}$ | Dimensionless time |

**Table 2** *Non-dimensional parameters used in the 2D model*

After the transformation, Eq. 7 and Eq. 2 reduce to:

$$w'''' + Pw'' - \alpha w'' \int_0^1 w'^2 dx + \ddot{w} = (\kappa_0 + \kappa_s + \kappa_d)\frac{1}{1+w} \quad (8)$$

where prime (') and dot (.) represent the derivative with respect to x and time, respectively. Doubly clamped boundary conditions are imposed: w(0,t)=w(1,t)=w'(0,t)=w'(1,t)=0. We describe the non-dimensional deflection as a superposition of the initial buckling ($w_0$), the static deflection ($w_s$) due to the DC voltage, and the dynamic oscillation ($w_d$) due to the AC actuation (Fig. 1):

$$w(x,t) = w_0(x) + w_s(x) + w_d(x,t) \tag{9}$$

Substituting Eq. 9 into Eq. 8, the CNT motion is described by a set of three nonlinear integro-differential equations: initial conditions (Eq. 10), static deflection due to the DC force, $\kappa_s$ (Eq. 11), and the dynamic motion due to the AC actuation $\kappa_d$ (Eq. 12):

$$\left[ w_0'''' + P w_0'' - \alpha w_0'' \int_0^1 w_0'^2 dx \right] \cdot (1 + w_0) = \kappa_0 \tag{10}$$

$$\left[ w_s'''' + \left( P - \alpha \int_0^1 w_0'^2 dx \right) w_s'' - \alpha \left( w_0'' + w_s'' \right) \int_0^1 \left( 2 w_0' w_s' + w_s'^2 \right) dx \right] \cdot (1 + w_0 + w_s) +$$
$$\left[ w_0'''' + P w_0'' - \alpha w_0'' \int_0^1 w_0'^2 dx \right] \cdot w_s = \kappa_s \tag{11}$$

$$\begin{bmatrix} w_d'''' + \left( P - \alpha \int_0^1 \left( w_0' + w_s' \right)^2 dx \right) w_d'' \\ -\alpha \left( w_0'' + w_s'' + w_d'' \right) \int_0^1 \left( 2 \left( w_0' + w_s' \right) w_d' + w_d'^2 \right) dx + \ddot{w}_d \end{bmatrix} \cdot (1 + w_0 + w_s + w_d) +$$
$$\left[ w_0'''' + w_s'''' + \left( P - \alpha \int_0^1 \left( w_0' + w_s' \right)^2 \right) \left( w_0'' + w_s'' \right) \right] \cdot w_d = \kappa_d \tag{12}$$

## 3. Galerkin Method

We approximate the beam deflection as:

$$w(x,t) = \sum_{i=1}^{N} \left( q_{0i} + q_{si} + q_{di}(t) \right) \varphi_i(x) \tag{13}$$

where $\varphi_i(x)$ is the i-th eigenmode of an unactuated doubly-clamped straight beam, given by:

$$\varphi_i(x) = \beta_i \left( \frac{\cos(\lambda_i) - 1}{\sin(\lambda_i) - \lambda_i} \sin(\lambda_i x) - \cos(\lambda_i x) + \lambda_i \frac{1 - \cos(\lambda_i)}{\sin(\lambda_i) - \lambda_i} x + 1 \right) \tag{14}$$

where $\beta_i$ are chosen such that $\max_{x=[0,1]}(\varphi_i(x))=1$ and $\lambda_i$ are the eigenvalues, found as the solution to $2(1 - \cos(\lambda_i)) - \lambda_i \sin(\lambda_i) = 0$. We shall note that $\varphi_i(x)$ with odd indices are symmetric functions of x in [0,1], whereas $\varphi_i(x)$ with even indices are asymmetric functions. The different $q_{0i}$, $q_{si}$, $q_{di}$ are the initial, static, and dynamic non-dimensional amplitudes of the $i^{th}$ eigenmode, respectively.

Assuming that $w_0(x) = \sum_{i=1}^{N} q_{0i} \varphi_i(x)$ is a solution to Eq. 10, we can substitute it into Eq. 11 as well as $w_s(x) = \sum_{i=1}^{N} q_{si} \varphi_i(x)$. Using the standard Galerkin decomposition[29,25], of multiplying the equation by $\varphi_i(x)$ and integrating over x within [0,1] interval, Eq. 11 is transferred to a system of coupled nonlinear algebraic equations, as detailed in the following sections.

For compact writing, let us define: $\int_0^1 \varphi_i dx = e_i$, $\int_0^1 \varphi_i^2 dx = a_{ii}$, $\int_0^1 \varphi_i'^2 dx = b_{ii}$, $\int_0^1 \varphi_i'''' \varphi_i dx = d_{ii}$, $\int_0^1 \varphi_i \varphi_j'''' \varphi_k dx = u_{ijk}$, $\int_0^1 \varphi_i \varphi_j'' \varphi_k dx = p_{ijk}$, $\int_0^1 \varphi_i \varphi_j \varphi_k dx = x_{ijk}$.

*3.1 Single modal*

In order to understand the significance of each modal to the overall behavior, let us begin taking only the first Galerkin modal. In this case, all equations can be solved analytically. The initial conditions are described by:

$$\left(d_{11} - \left(P - \alpha b_{11} q_{01}^2\right) b_{11}\right)\left(1 + q_{01}\right) = \kappa_0 \qquad (15)$$

For $\kappa_0 = 0$ we can extract a simple relation between the initial axial strain and the initial midpoint deflection: $q_{01}^2 = (Pb_{11} - d_{11}) / (\alpha b_{11}^2)$, where we choose the negative root for upward buckling and the positive root for downward buckling. In order to receive some initial buckling (i.e. a real root) we must demand $P > d_{11}/b_{11}$, which translates to the standard EB buckling criterion $|T_0| > 4\pi^2 EI / L^2$. However, if we allow $\kappa_s \neq 0$ then P and $q_{01}$ can be chosen as two independent parameters, yielding more diverse possibilities for the static and dynamic behavior of the device.

Assuming independent P and $q_{01}$, and taking only one modal (N=1), one receives a cubic equation for $q_{s1}$ for every static load $\kappa_s$:

$$q_{s1}\left(d_{11} - \left(P - \alpha\left(Q_0 + Q_s\right)\right)b_{11}\right)\left(1 + \lambda 2 q_{01} + \lambda q_{s1}\right) + \alpha q_{01} b_{11} Q_s \left(1 + \lambda q_{01}\right) = \kappa_s e_1 \qquad (16)$$

where $Q_0 = b_{11} q_{01}^2$, $Q_s = 2 b_{11} q_{01} q_{s1} + q_{s1}^2 b_{11}$, and $\lambda = 1$ (its role will be clarified in the next paragraph).

As a sanity check, one can notice that for $\kappa_s = 0$ we get the trivial solution, that corresponds to our definition of the static movement: $q_{s1} = 0$.

### 3.1.1 Single modal with constant force

The static equation (16) defines the dependence of the static deflection of the beam, $q_{s1}$, on the applied static force, $\kappa_s$, but it is customary to visualize the static response as $\kappa_s$ vs. $q_{s1}$ (as in Fig. 2) since the force can also be regarded as the derivative of the potential energy ($\kappa_s = dU/dq_{s1}$). Therefore, physical solutions (i.e. stable solutions) exist only for non-negative values of the derivative of Eq. 16, meaning $d^2U/dq_{s1}^2 \geq 0$. Snap-through transition occurs at the extremum points of this curve, where $d\kappa_s/dq_{s1}=0 \rightarrow d^2U/dq_{s1}^2=0$. Hence, if one wishes to find the critical loads at which the ST transition occurs, one should compare the derivative of Eq. 16 to zero.

If we assume small initial and static deflection ($\hat{w} \ll g_0$), we can approximate the static electrostatic force as a constant: $\hat{\kappa}_s \approx \left(\pi\varepsilon_0 LV_{gDC}^2\right)/\left(g_0 \ln^2(2g_0/r)\right)$. Substituting into Eq. 1, transforming the equation to dimensionless units, and performing the Galerkin decomposition process for a single modal (N=1), we obtain the same static equation as in Eq. 16, but with $\lambda = 0$:

$$q_{s1}\left(d_{11} - \left(P - \alpha(Q_0 + Q_s)\right)b_{11}\right) + \alpha q_{01} b_{11} Q_s = \kappa_s e_1 \tag{17}$$

Substituting $Q_0$ and $Q_s$, taking the derivative of Eq. 17 with respect to $q_{s1}$, and comparing to zero yields

$$q_{s1}^{extremum} = \left(-6q_{01} \pm \sqrt{-12(d_{11} - Pb_{11})/\alpha/b_{11}^2}\right)/6 \tag{18}$$

From here we find the non-dimensional criterion for buckling $P_c = d_{11}/b_{11}$, as in previous studies[30,31]. In the case where $P<P_c$, the roots are imaginary and hence the transition from upward to downward curvature is continuous, but for $P>P_c$, there exist two real extremum points in which the snap-through and release transitions occur (marked by the dashed arrows in Fig. 2). If we were to assume that $\kappa_0=0$ and substitute the condition it imposes on $q_{01}$ into Eq. 16, the continuous transition is not possible. Furthermore, bi-stability in this case ($\kappa_0=0$) results only in latching[17], meaning a snap-back (release) transition at $\kappa_s<0$.

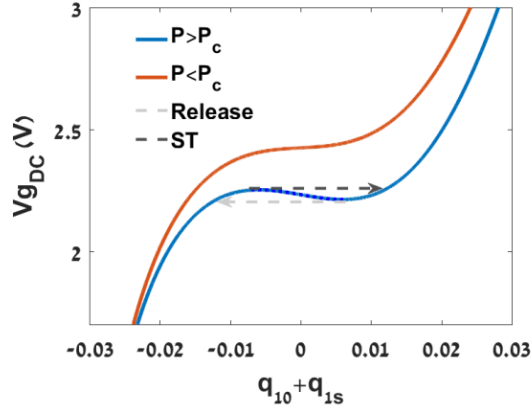

**Figure 2 Single modal static response.** *Bifurcation diagram of the single modal approximation with a constant force. When $P<P_c$ (orange) the transition from upward ($q_{01}+q_{s1}>0$) to downward ($q_{01}+q_{s1}>0$) buckling is continuous, and when $P>P_c$ (blue) an unstable solution is formed (dashed part of the blue curve) and the transitions are snap-through buckling and release, marked by the two arrows.*

3.1.2. General solution

Realistic device parameters force us to consider the nonlinearity of the electrostatic force. Unfortunately, an analytical solution is not possible. Therefore, to investigate the system, we introduce the nondimensional parameter $\lambda$ to Eq. 16, where $\lambda = 0$ reduces the equation to the case of constant force (Eq. 17) and $\lambda = 1$ is the general equation (Eq. 16):

$$q_{s1}\left(d_{11}-(P-\alpha(Q_0+Q_s))b_{11}\right)(1+2\lambda q_{01}+\lambda q_{s1})+\alpha q_{01}b_{11}Q_s(1+\lambda q_{01})=\kappa_s e_1 \qquad (19)$$

Examining the effect of $\lambda$ reveals that the qualitative behavior does not change, only that a smaller force is required for ST and release (Fig. 3).

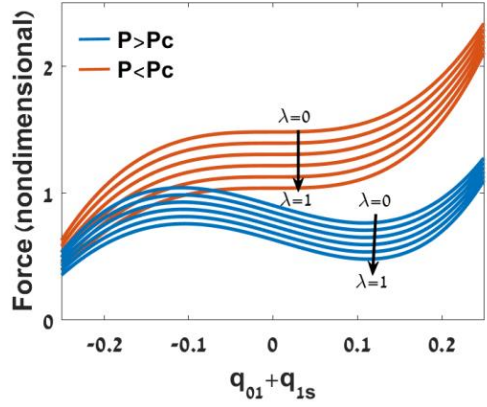

**Figure 3 Effect of nonlinear force.** *Bifurcation diagram of the single modal approximation, taking the electrostatic nonlinearity into account ($\lambda=1$). The transition from upward to downward buckling, whether continuous ($P<P_c$) or through ST ($P>P_c$), occurs at a smaller static load than the solution to Eq. 17.*

3.1.3. Single modal eigenvalue problem analysis

Moving to the dynamic equation, we perform the Galerkin decomposition for Eq. 12, yielding:

$$\begin{aligned}&\left[q_{d1}d_{11}-\left(P-\alpha(Q_0+Q_s)\right)q_{d1}b_{11}+\alpha(q_{01}+q_{s1}+q_{d1})b_{11}Q_d\right]\cdot(1+q_{01}+q_{s1}+q_{d1})\\&+\left[(q_{01}+q_{s1})d_{11}-\left(P-\alpha(Q_0+Q_s)\right)(q_{01}+q_{s1})b_{11}\right]\cdot q_{d1}\\&-\left[q_{d1}a_{11}-(q_{01}+q_{s1}+q_{d1})x_{111}\right]\cdot\omega^2=\kappa_d e_1\end{aligned} \quad (20)$$

where $Q_d = 2(q_{01}+q_{s1})q_{d1}b_{11}+q_{s1}^2 b_{11}$. Assuming small vibrations, we can neglect the nonlinear dynamic terms and solve for the resonance frequencies. As displayed in Fig. 4a, if $P < P_c$ the resonance frequency dependence on the gate voltage is continuous, whereas for $P > P_c$ the resonance frequency decreases to zero after which it exhibits a "jump". Figs. 4b-c present the corresponding static behavior for the continuous transition (b) vs a ST mechanical "jump" (c). As both of the qualitative behaviors in Fig. 4a are observed in our devices[21], this result is very encouraging. However, the "jumps" measured in our experiments occur at non-zero resonance frequency, and we did not detect zero resonance frequency in any of the dozens of fabricated devices.

Previous studies[30,31] suggest that symmetry breaking can account for such observations. In order to introduce such symmetry breaking into the system we must take two or more modals into our Galerkin based approximation.

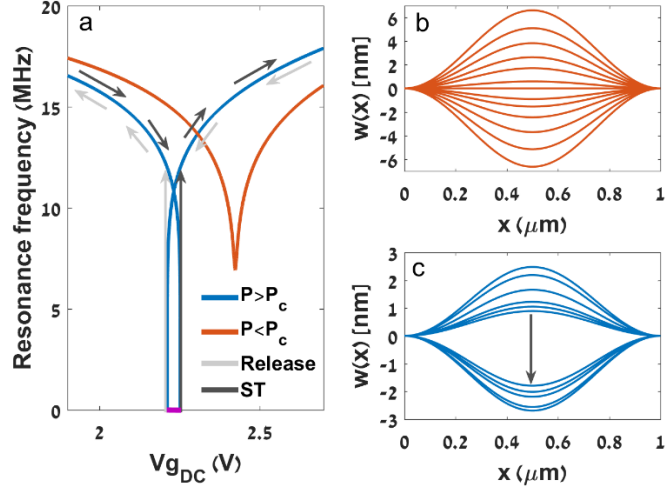

**Figure 4 Single modal dynamic response.** *(a) Resonance frequency dependence on the static gate voltage for a single modal approximation for both cases where $P<P_c$ (continuous transition, orange) and $P>P_c$ (exhibiting a "jump", blue). The dark and light grey arrows present an upward or downward gate voltage sweeps, respectively. Purple line represents the ST hysteresis window, i.e. the difference between static loads at which the ST (dark grey) and release (light grey) jumps occur. (b-c) CNT shape for several static loads in the vicinity of the transition from upward to downward buckling, corresponding to the curves in (a), presenting both continuous transition for $P<P_c$ (b) and ST "jump" (c).*

*3.2 Two modals*

3.2.1. Static response
In previous studies, even when a second modal was introduced, the initial shape still consisted of only a single modal, $q_{01}$[30]. This makes sense for a buckled micro-beam since the width is typically much bigger than the thickness. However, for a naturally-grown CNT, the width and thickness are the same, and the chances of the CNT to grow in a perfect symmetric arch shape are quite slim. In addition, we show that if the deflection is very small such that the electrostatic nonlinearity can be neglected, taking $q_{02}=0$ in Eq. 17 eliminates the geometric nonlinearity and results in a linear dependence of $\kappa_s$ on $q_{s1}$ (see supplementary information for details). Therefore, we must introduce a second modal to the initial conditions as well.

After the Galerkin analysis for two modals, a solution to Eq. 11 can be approximated by solving the following set of coupled equations:

$$\left[q_{s1}d_{11}-(P-\alpha Q_0)q_{s1}b_{11}+\alpha b_{11}(q_{01}+q_{s1})\cdot Q_s\right]\cdot(1+q_{01}+q_{s1})+$$
$$\left[q_{s2}u_{221}+(P-\alpha Q_0)q_{s2}p_{221}-\alpha p_{221}(q_{02}+q_{s2})\cdot Q_s\right]\cdot(q_{02}+q_{s2})+$$
$$+q_{s1}\left[q_{01}d_{11}-(P-\alpha Q_0)b_{11}q_{01}\right]+q_{s2}\left[q_{02}u_{221}+(P-\alpha Q_0)p_{221}q_{02}\right]-\kappa_s e_1=0$$

$$\left[q_{s2}d_{22}-(P-\alpha Q_0)b_{22}q_{s2}+\alpha b_{22}(q_{02}+q_{s2})\cdot Q_s\right]+$$
$$\left[q_{s2}u_{221}+(P-\alpha Q_0)p_{221}q_{s2}-\alpha p_{221}(q_{02}+q_{2s})\cdot Q_s\right]\cdot(q_{01}+q_{s1})+$$
$$\left[q_{s1}u_{212}+(P-\alpha Q_0)p_{212}q_{1s}-\alpha p_{212}(q_{01}+q_{s1})\cdot Q_s\right]\cdot(q_{02}+q_{s2})+$$
$$q_{s1}\left[q_{02}u_{221}+(P-\alpha Q_0)p_{221}q_{02}\right]+q_{s2}\left[q_{01}u_{212}+(P-\alpha Q_0)p_{212}q_{01}\right]=0$$

(21)

where $Q_0 = q_{01}^2 b_{11}+q_{02}^2 b_{22}$ and $Q_s = b_{11}q_{s1}(2q_{01}+q_{s1})+b_{22}q_{s2}(2q_{02}+q_{s2})$.

Before, the only parameter responsible for the formation of snap-through buckling was the initial axial strain P. In the two modals case, however, the initial $q_{02}$ also significantly influences the static response of the system, and whether a snap-through or a continuous transition will occur. Fig. 5 presents the solutions of Eq. 21 for the force dependence on the symmetric static modal ($q_{s1}$), solved for varying asymmetric initial conditions (i.e. varying $q_{02}$ values). It can be observed that when $q_{02}$ is small, the stable solution (thick line) is not continuous, and when $q_{02}$ increases, the stable solution becomes continuous (green line). Note that even when $q_{02} \rightarrow 0$, while the mathematical solution (blue lines) approaches the single modal solution (grey line), the physically stable part (thick blue line in Fig. 5) is different, and the ST and release transitions occur at a new extremum point of the $\kappa_s$ vs. $q_{s1}$ curve (at the edges of the thick lines, marked by the light blue dots), which is far from the single-modal solution extrema (yellow dots, grey line).

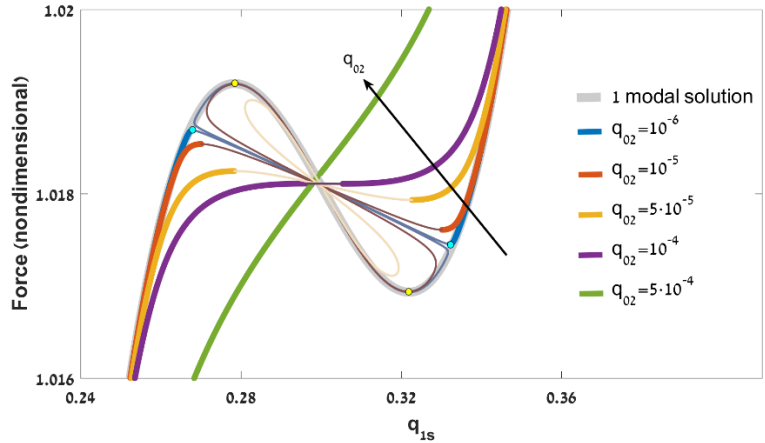

**Figure 5 Two modals static response.** *The dependence of the static response on the initial second modal $q_{02}$. The different colored lines represent different initial $q_{02}$ values, where the thick parts are stable solutions and the thin lines represent an unstable solution. The grey line is the solution for a single modal approximation, in which $P>P_c$. $q_{02}\rightarrow 0$ approaches the single modal solution, but the ST and release transitions (light blue dots) occur before reaching the extremum points of the grey curve (yellow dots). Larger $q_{02}$ results in a continuous transition.*

Let us explain. The first equation in Eq. 21 defines a two-dimensional surface of the force, $\kappa_s(q_{s1},q_{s2})$ (colored surface in Fig. 6). The left column presents a three-dimensional view of the force for increasing $q_{02}$ values. Together with the second equation, $\kappa_s(q_{s1},q_{s2}(q_{s1}))$ is reduced to a curve in this three-dimensional space, depicted by the black (stable) and red (unstable) lines in Fig. 6. The middle column presents a "front view" of the same graph as in the left column, perpendicular to the $\kappa_s$-$q_{s1}$ plane, such that the edge of the colored surface projection is actually the single-modal solution, showcasing how increasing $q_{02}$ shifts the 2-modal solution (black and blue curve) away from the single-modal curve (as in Fig. 5). The addition of the asymmetric mode creates a "bypass" around the single modal bifurcation points through the $q_{s2}$ axis, as can be observed in the right column of Fig. 6, which is a "side-view" of the same graph as the left column, almost perpendicular to the $\kappa_s$-$q_{s2}$ plane. The snap-through transition occurs only when reaching a local extremum point along the $\kappa_s(q_{s1},q_{s2}(q_{s1}))$ curve, where $d\kappa_s/dq_{s1} = \partial\kappa_s/\partial q_{s1} + \partial\kappa_s/\partial q_{s2}\cdot \partial q_{s2}/\partial q_{s1} = 0$ (exactly at the edges of the black curve in Fig. 6 mid-column or the edges of the thick colored lines in Fig. 5). When $q_{02}$ increases, while the landscape of $\kappa_s$ is only slightly modified, the solution curve moves away from the saddle point along the $q_{s2}$ axis, bypassing the instability. In the mechanical system, what actually happens to the CNT is referred to as symmetry breaking. The CNT shape can now deform before the transition from upward to downward buckling, both for the ST jump (Fig. 7b) as well as for the continuous transition (Fig. 7a). Since $q_{01}>>q_{02}$ the initial beam shape is almost symmetric. When approaching the transition point, however, the symmetric mode nearly vanishes and the beam shape transforms to be nearly asymmetric. This corresponds well with the shift along the $q_{s2}$ axis, apparent in the right column of Fig. 6 (see also Fig. S1).

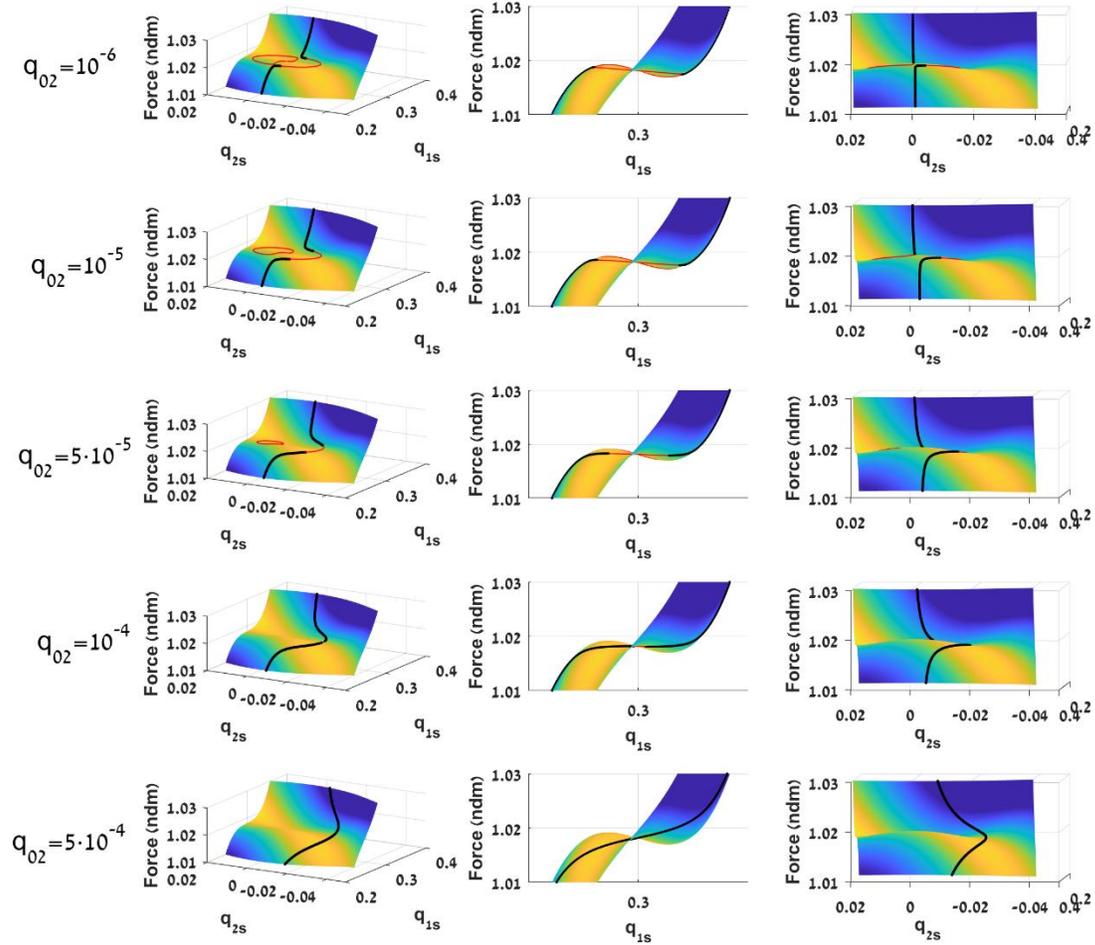

**Figure 6 Effect of initial conditions.** *The dependence of the static response on the initial second modal $q_{02}$. As $q_{02}$ increases (top to bottom), the force curve is shifted from the saddle point and the bifurcation is avoided. Left column is a three-dimensional view. Middle column is the $\kappa$-$q_{s1}$ cross-section (same as in Fig. 5) and right column is the $\kappa$-$q_{s2}$ cross-section. The black and red line indicates the $\kappa_s(q_{s1},q_{s2}(q_{s1}))$ curve in this three-dimensional space, that is the solution to Eqs. 21, where black parts represent stable solution and red parts represent unstable solutions.*

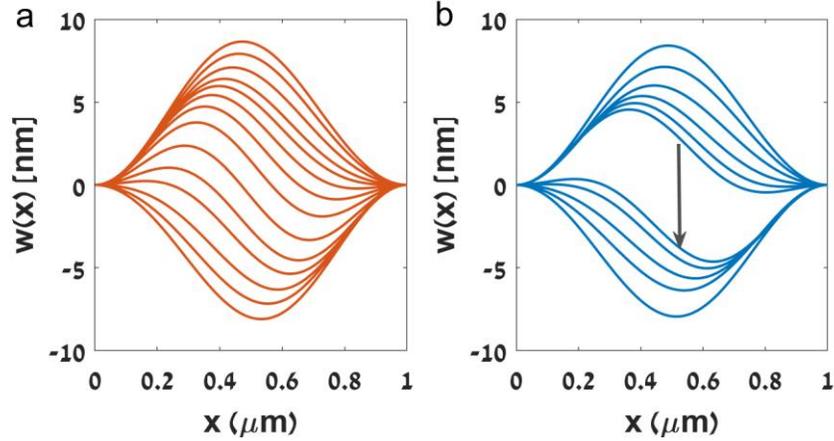

**Figure 7 Symmetry breaking.** *CNT shape for several static loads in the vicinity of the transition from upward buckling to downward buckling, both for continuous transition (a), corresponding to $q_{02}=2e\text{-}4$, as well as ST transition (b), corresponding to $q_{02}=5e\text{-}5$. Symmetry breaking is observed for both cases.*

### 3.2.2. Eigenvalue problem analysis

Moving to the dynamics, we perform the Galerkin decomposition process for Eq. 12, which transforms into:

$$(q_{01}+q_{s1})q_{d1}u_{111}+(q_{02}+q_{s2})q_{d2}u_{221}+q_{d1}d_{11}$$
$$+(q_{01}+q_{s1}+q_{d1})q_{d1}u_{111}+(q_{02}+q_{s2}+q_{d2})q_{d2}u_{221}+$$
$$(P-\alpha Q_{0d})\cdot\begin{bmatrix}(q_{01}+q_{s1})q_{d1}p_{111}+(q_{02}+q_{s2})q_{d2}p_{221}-q_{d1}b_{11}\\+(q_{01}+q_{s1}+q_{d1})q_{d1}p_{111}+(q_{02}+q_{s2}+q_{d2})q_{d2}p_{221}\end{bmatrix}-$$
$$\alpha Q_d\cdot\left[-(q_{01}+q_{s1}+q_{d1})b_{11}+(q_{01}+q_{s1}+q_{d1})^2 p_{111}+(q_{02}+q_{s2}+q_{d2})^2 p_{221}\right]-$$
$$\omega^2\cdot\left[q_{d1}a_{11}+(q_{01}+q_{s1}+q_{d1})q_{d1}x_{111}+(q_{02}+q_{s2}+q_{d2})q_{d2}x_{221}\right]-e_1\kappa_d=0$$

(21)

$$(q_{02}+q_{s2})q_{d1}u_{122}+(q_{01}+q_{s1})q_{d2}u_{212}+q_{d2}d_{22}$$
$$+(q_{01}+q_{s1}+q_{d1})q_{d2}u_{122}+(q_{02}+q_{s2}+q_{d2})q_{d1}u_{212}+$$
$$(P-\alpha Q_{0d})\cdot\begin{bmatrix}(q_{02}+q_{s2})q_{d1}p_{122}+(q_{01}+q_{s1})q_{d2}p_{212}-q_{d2}b_{22}\\+(q_{01}+q_{s1}+q_{d1})q_{d2}p_{122}+(q_{02}+q_{s2}+q_{d2})q_{d1}p_{212}\end{bmatrix}-$$
$$\alpha Q_d\cdot\left[-(q_{02}+q_{s2}+q_{d2})b_{22}+(q_{01}+q_{s1}+q_{d1})(q_{02}+q_{s2}+q_{d2})(p_{122}-p_{212})\right]-$$
$$\omega^2\cdot\left[q_{d2}a_{22}+(q_{01}+q_{s1}+q_{d1})q_{d2}x_{212}+(q_{02}+q_{s2}+q_{d2})q_{d1}x_{122}\right]=0$$

where $Q_{0d}=\int\left(w_0'^2+2w_0'w_s'+w_s'^2\right)dx=(q_{01}+q_1)^2 b_{11}+(q_{02}+q_2)^2 b_{22}$ and

$$Q_d = \int \left(2w_0' w_d' + 2w_s' w_d' + w_d'^2\right) dx = 2b_{11}(q_{01}+q_{s1})q_{d1} + 2b_{22}(q_{02}+q_{s2})q_{d2} + q_{d1}^2 b_{11} + q_{d2}^2 b_{22}.$$

We wish to solve for the resonance frequencies, so we neglect all nonlinear terms. As this is essentially an eigenvalue problem, it can be described as the following:

$$\begin{pmatrix} q_{d1} & q_{d2} \end{pmatrix} \begin{pmatrix} H_{11} & H_{12} \\ H_{21} & H_{22} \end{pmatrix} \begin{pmatrix} q_{d1} \\ q_{d2} \end{pmatrix} = \omega^2 \begin{pmatrix} q_{d1} \\ q_{d2} \end{pmatrix} \quad (22)$$

where $H_{11}=\partial^2 H/\partial q_{d1}^2$, $H_{22}=\partial^2 H/\partial q_{d2}^2$, $H_{12}=\partial^2 H/\partial q_{d1} \partial q_{d2}$, $H_{21}=\partial^2 H/\partial q_{d2} \partial q_{d1}$, and H is the system's Hamiltonian. The eigenvalues are found from the requirement that the determinant of H is zero: $\text{Det}(H)=(H_{11}-\omega^2)(H_{22}-\omega^2)-H_{21}H_{12}=0$. The snap-through buckling is a critical bifurcation point, where the Jacobian $J=H_{11}H_{22}-H_{12}H_{21}$ also vanishes. Therefore, at the critical point, the eigenvalue problem reduces to:

$$\omega^2(\omega^2-(H_{11}-H_{22})) = 0 \quad (23)$$

which has two real and non-negative solutions: $\omega_1=0$ and $\omega_2=(|H_{11}-H_{22}|)^{1/2}$, corresponding to the first and second resonance modes. This means that at the unique snap-through bifurcation point there must be a resonance mode approaching zero frequency. This analysis is holds for any number of Galerkin modals, the conditions of $\text{Det}(H)=0$ and $J=0$ will always impose at least one zero frequency mode at the bifurcation point. Fig. 8 presents the eigenvalue problem solution to Eq. 21. It depicts the first resonance mode dependence on the static force for the case of a small asymmetric initial component (blue), which results in a snap-through transition vs. the case of a large asymmetric initial components (orange), which results in a continuous transition. For small $q_{02}$ (blue curve), the resonance frequency decreases due to compression (dark grey arrows) until it reaches zero, and then the snap-through transition occurs, which translates to a "jump" in the resonance frequency (marked by the vertical dark grey arrow). Increasing the force further (dark grey arrows) causes the resonance frequency to rise due to stretching (hardening). If the force is slowly reduced back (i.e., gate voltage is swept back downward) the resonance frequency decreases along the right-hand blue curve (light grey arrows) until reaching zero again, but for a smaller force, resulting in a snap-back (release) transition (marked by the vertical light grey arrow) to an upward configuration. The difference between the gate voltages at which the snap-through and release transitions occurs creates a hysteresis window (purple double-arrow). For larger $q_{02}$ (orange) the resonance frequency decreases as before, until reaching a non-zero minimum, after which it increases, corresponding to a continuous mechanical transition from upward to downward curvature, i.e., from compression to stretching without a "jump".

Unfortunately, as explained at the end of Sect. 3.1, in our experimental data[21], the "jump" in frequency as a result of the ST transition occurs at nonzero frequency, and hence the addition of a second modal did not solve this issue. The only way to solve this puzzle is the addition of out-of-plane motion, which shall be discussed in section 4.

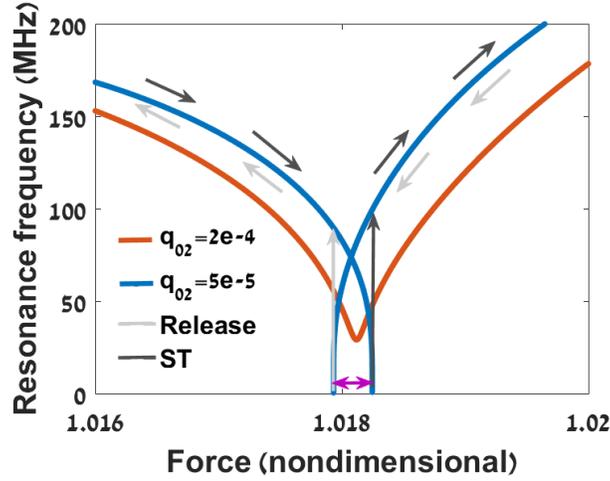

**Figure 8 Two modals dynamic response.** *Resonance dependence on the static load for a two-modal approximation, where $P>P_c$. The existence of a snap-through transition ("jump") is dependent on the initial $q_{02}$. Blue curve present a solution for small $q_{02}$ in which ST transition occurs, and orange curve presents the solution for larger $q_{02}$ which enables continuous transition. The dark and light grey arrows present an upward or downward gate voltage sweeps, respectively. Symmetry breaking is evident even for a small $q_{02}$ (blue), as the ST and release "jumps" (vertical arrows) are not equal. Purple double-arrow represents the ST hysteresis window, i.e. the difference between static loads at which the ST (dark grey) and release (light grey) jumps occur.*

### 3.3 Finite Element Method

To verify our results from the reduced order model, we solved Eq. 7 using finite element method as well[32]. The transition to a non-dimensional equation is based again on Table 2, except that in this case we do **not** assume $ln\left(\frac{2(g_0+\hat{w})}{r}\right) \approx ln\left(\frac{g_0}{r}\right)$, and take the complete expression for the force. We divide the beam by n grid points to n+1 finite segments. We use the cubic (Hermite) approximation for the displacement of each segment[33]:

$$w(x,t) = \sum_{i=1}^{n}\left(p_i(t)\phi_i(x) + q_i(t)\psi_i(x)\right) \qquad (24)$$

where h is the segment's length, such that $x_i=(i-1)h$, and $x_0=0$, $x_n=L$. We define: $p_i(t) = w_n(x_i,t)$ and $q_i(t) = h \cdot \frac{\partial w_n}{\partial x}(x_i,t)$ where

$$\phi_i(x) = \begin{cases} 1 - 3\xi_{i-1}^2 - 2\xi_{i-1}^3, & x_{i-1} \leq x \leq x_i \\ 1 - 3\xi_i^2 + 2\xi_i^3, & x_i \leq x \leq x_{i+1} \\ 0, & x \leq x_{i-1} \text{ and } x \geq x_{i+1} \end{cases}$$

$$\psi_i(x) = \begin{cases} h_{i-1} \cdot \left(\xi_{i-1} + 2\xi_{i-1}^2 + \xi_{i-1}^3\right), & x_{i-1} \leq x \leq x_i \\ h_i \cdot \left(\xi_i - 2\xi_i^2 + \xi_i^3\right), & x_i \leq x \leq x_{i+1} \\ 0, & x \leq x_{i-1} \text{ and } x \geq x_{i+1} \end{cases}$$

in which

$$\begin{cases} \xi_{i-1} = (x - x_{i-1})/h_{i-1}, & x_{i-1} \leq x \leq x_i \\ \xi_i = (x - x_i)/h_i, & x_i \leq x \leq x_{i+1} \end{cases}$$

and $h_{i-1}=x_i-x_{i-1}$, $h_i=x_{i+1}-x_i$. Doubly clamped boundary conditions are imposed: $p_i(t) = q_i(t) = 0$, $i \in \{0, n+1\}$. We substitute these expressions into the dimensionless static and dynamic equations, and receive a set of coupled algebraic equations for $p_i$, $q_i$, $\omega$. The initial conditions (initial beam shape) are chosen to be the same as before: $w_0 = q_{01}\varphi_1 + q_{02}\varphi_2$.

We compare the results obtained by solving the beam equation using the Galerkin approximation for two modals, with the FEM method for n=23 (Fig. 9), using the physical parameters detailed in Table 1 and initial conditions typical for the fitted devices. As the comparison between the FEM and Galerkin approximations yielded very similar results, we can conclude that the approximation regarding the force (the transition from Eq. 6 to Eq. 7) is justified, and that using only two modals in the Galerkin method is sufficient (meaning, higher orders can be neglected).

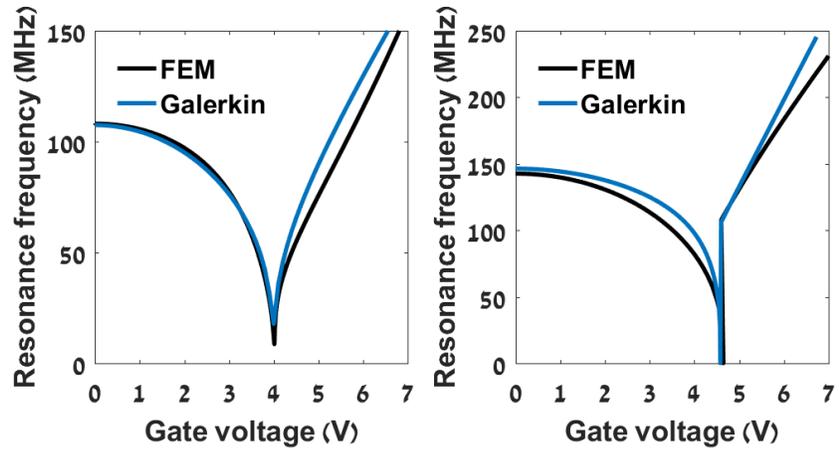

**Figure 9 Galerking vs. FEM.** *Comparison between the finite element method approximation and the Galerkin two modals approximation for continuous transition (left) and a snap-through transition (right).*

## 4. Three Dimensional Model

Unlike traditional micro beams, where the beam width is usually much larger than its thickness (b>>h in Ref. 26), the CNT cross section is circular and therefore symmetric. Hence, it is free to move in the y direction (namely, out-of-plane motion) just as in the z direction (in-plane motion). For a CNT with a downward slack ($q_{01}>0$) the static out-of-plane motion is less likely, since the torque attracts the CNT towards in-plane motion. However, for a CNT with $q_{01}<0$, the torque pushes the tube away from the in-plane motion, and the whole movement from upward curvature towards the gate electrode can include out-of-plane motion as well and cannot be ignored. Therefore, we are forced to describe the CNT motion as a superposition of in-plane ($w(x,t)$) and out-of-plane ($v(x,t)$) components:

$$w(x,t) = w_0(x) + w_s(x) + w_d(x,t) \tag{25}$$

$$v(x,t) = v_0(x) + v_s(x) + v_d(x,t) \tag{26}$$

In this case we receive two coupled non-dimensional Euler-Bernoulli beam equations for each component, where we still restrict the force exerted by the gate to be solely along the z direction (in-plane):

$$w'''' + Pw'' - \alpha w'' \int_0^1 \left(w'^2 + v'^2\right)dx + \ddot{w} = \left(\kappa_0 + \kappa_s + \kappa_d\right)\frac{1}{1+w} \tag{27}$$

$$v'''' + Pv'' - \alpha v'' \int_0^1 \left(w'^2 + v'^2\right)dx + \ddot{v} = 0 \tag{28}$$

As before, these can be divided into initial conditions, static equations and dynamic equations:

$$\begin{cases} \left[ w_0'''' + Pw_0'' - \alpha w_0'' Q_0 \right] \cdot (1 + w_0) = \kappa_0 \\ v_0'''' + Pv_0'' - \alpha v_0'' Q_0 = 0 \end{cases}$$

$$\begin{cases} \left[ w_s'''' + (P - \alpha Q_0) w_s'' - \alpha \left( w_0'' + w_s'' \right) Q_s \right] \cdot (1 + w_0 + w_s) + \left[ w_0'''' + Pw_0'' - \alpha w_0'' Q_0 \right] \cdot w_s = \kappa_s \\ v_s'''' + (P - \alpha Q_0) v_s'' - \alpha \left( v_0'' + v_s'' \right) Q_s = 0 \end{cases} \quad (29)$$

$$\begin{cases} \left[ w_d'''' + (P - \alpha (Q_0 + Q_s)) w_d'' - \alpha \left( w_0'' + w_s'' + w_d'' \right) Q_d + \ddot{w}_d \right] \cdot (1 + w_0 + w_s + w_d) \\ \quad + \left[ w_0'''' + w_s'''' + (P - \alpha (Q_0 + Q_s))(w_0'' + w_s'') \right] \cdot w_d = \kappa_d \\ v_d'''' + (P - \alpha (Q_0 + Q_s)) v_d'' - \alpha \left( v_0'' + v_s'' + v_d'' \right) Q_d + \ddot{v}_d = 0 \end{cases}$$

where the in-plane and out-of-plane coupling results from the built-in tension along the CNT, represented by the following integral terms:

$$\begin{aligned} Q_0 &\triangleq \int_0^1 \left( w_0'^2 + v_0'^2 \right) dx \\ Q_s &\triangleq \int_0^1 \left( 2w_0' w_s' + w_s'^2 + 2v_0' v_s' + v_s'^2 \right) dx \\ Q_d &\triangleq \int_0^1 \left( 2\left( w_0' + w_s' \right) w_d' + w_d'^2 + 2\left( v_0' + v_s' \right) v_d' + v_d'^2 \right) dx \end{aligned} \quad (30)$$

*4.1 Static analysis*

After verifying that two modals provide a satisfying approximation, we substitute

$$\begin{aligned} w_0(x) &= q_{01} \varphi_1(x) + q_{02} \varphi_2(x) \\ w_s(x) &= q_{s1} \varphi_1(x) + q_{s2} \varphi_2(x) \\ v_0(x) &= v_{01} \varphi_1(x) + v_{02} \varphi_2(x) \\ v_s(x) &= v_{s1} \varphi_1(x) + v_{s2} \varphi_2(x) \end{aligned} \quad (31)$$

into Eqs. 29 and receive a set of four coupled algebraic static equations:

$$\left[q_{s1}d_{11} - (P-\alpha Q_0)q_{s1}b_{11} + \alpha b_{11}(q_{01}+q_{s1})\cdot Q_s\right]\cdot(1+q_{01}+q_{s1})+$$
$$\left[q_{s2}u_{221} + (P-\alpha Q_0)q_{s2}p_{221} - \alpha p_{221}(q_{02}+q_{s2})\cdot Q_s\right]\cdot(q_{02}+q_{s2})+$$
$$+q_{s1}\left[q_{01}d_{11} - (P-\alpha Q_0)b_{11}q_{01}\right] + q_{s2}\left[q_{02}u_{221} + (P-\alpha Q_0)p_{221}q_{02}\right] - \kappa_s e_1 = 0$$

$$\left[q_{s2}d_{22} - (P-\alpha Q_0)b_{22}q_{s2} + \alpha b_{22}(q_{02}+q_{s2})\cdot Q_s\right]+$$
$$\left[q_{s2}u_{221} + (P-\alpha Q_0)p_{221}q_{s2} - \alpha p_{221}(q_{02}+q_{2s})\cdot Q_s\right]\cdot(q_{01}+q_{s1})+ \quad (32)$$
$$\left[q_{s1}u_{212} + (P-\alpha Q_0)p_{212}q_{1s} - \alpha p_{212}(q_{01}+q_{s1})\cdot Q_s\right]\cdot(q_{02}+q_{s2})+$$
$$q_{s1}\left[q_{02}u_{221} + (P-\alpha Q_0)p_{221}q_{02}\right] + q_{s2}\left[q_{01}u_{212} + (P-\alpha Q_0)p_{212}q_{01}\right] = 0$$

$$v_{s1}d_{11} - (P-\alpha Q_0)v_{s1}b_{11} + \alpha(v_{01}+v_{s1})b_{11}\cdot Q_s = 0$$

$$v_{s2}d_{22} - (P-\alpha Q_0)v_{s2}b_{22} + \alpha(v_{02}+v_{s2})b_{22}\cdot Q_s = 0$$

where $Q_0 = b_{11}(q_{01}^2+v_{01}^2) + b_{22}(q_{02}^2+v_{02}^2)$ and $Q_s = b_{11}(q_{s1}^2+v_{s1}^2) + b_{22}(q_{s2}^2+v_{s2}^2) + 2b_{11}(q_{01}q_{s1}+v_{01}v_{s1}) + 2b_{22}(q_{02}q_{s2}+v_{02}v_{s2})$. Note that the in-plane equations remain the same as Eqs. 21, and the coupling to the out-of-plane motion is manifested through the built-in tension integrals, $Q_0$ and $Q_s$.

Equations 32 can be solved numerically. Fig. 10 presents the effect of the initial out-of-plane component $v_{01}$ on the static response of the beam; $q_{02}$=-1e-5 was chosen such that if $v_{01}$=0, 2D ST should occur. $v_{02}$ is negligible for all curves. For higher values of $v_{01}$, but even for very small values, we see that the snap-through transition is eliminated, and the CNT moves in a continuous torsional motion downwards (Fig. 11). The effect of $v_{02}$ is very similar to that of $q_{02}$ in Fig. 5. Fig. 11a presents a three dimensional rotational transition of the static CNT shape from upward to downward curvature as the voltage applied to the LG is increased, for the case where $v_{01}$=-1e-5. Bottom-left panel (Fig. 11b) is a "side-view" (projection on the zy-plane), showing how the out-of-plane rotational motion evolves to avoid ST buckling, and the bottom-right panel (Fig. 11c) is a "front-view" (projection on the xz-plane), showcasing a continuous transition; no "jump" is observed.

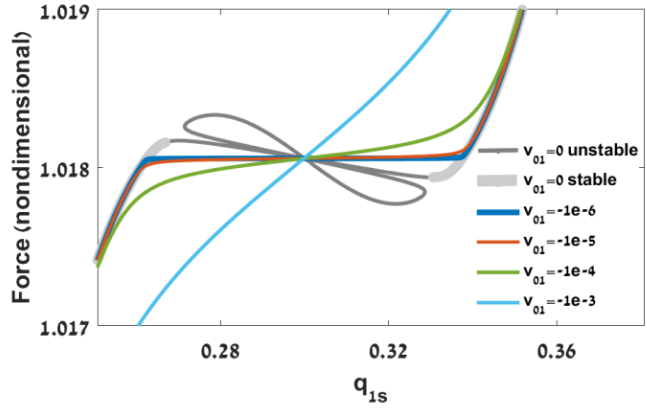

**Figure 10 Three dimensional static response.** *The dependence of the static response on the initial first out-of-plane modal $v_{01}$. Even at very small $v_{01}$ values, the stable solution is continuous. Raising $v_{01}$ increases the slope of the continuous curve.*

We shall note a significant difference between the in-plane and out-of-plane static deflection. Fig. 12 depicts two scenarios extracted from the model. For initial conditions in which $v_{01,02} \ll q_{01}$ the evolution of the out-of-plane static modals ($v_{s1,s2}$) is plotted in Figs. 12a and Fig. 12b. Fig. 12a presents their amplitudes with respect to the mid-point z-deflection ($q_{01} + q_{s1}$), and Fig. 12b depicts their relative ratio ($v_{s1}/v_{s2}$). One can observe that indeed at the ST transition the two components reach their maximum values with a 10 percent increase of the symmetric modal ($v_{s1}$) compared to the anti-symmetric one ($v_{s2}$). However, for higher values of the out-of-plane initial conditions ($v_{01,02} \leq q_{01}$), out-of-plane centering of the CNT shape is observed (Fig. 12c,d). Unlike the in-plane symmetry breaking in which near the transition, the in-plane symmetric mode, $q_{s1}$, vanishes and the asymmetric mode, $q_{s2}$, is at its maximum (Fig. 7 and Fig. S1), the out-of-plane motion act in an opposite manner. Near the transition, when the out-of-plane deflection is at its maximum, the asymmetric component, $v_{s2}$, is oppressed compared to the symmetric component, $v_{s1}$ (Fig. 12c,d). Moreover, since the tube length increases due to stretching by the LG, the maximum out-of-plane deflection occurs at positive z ($q_{01} + q_{s1} > 0$).

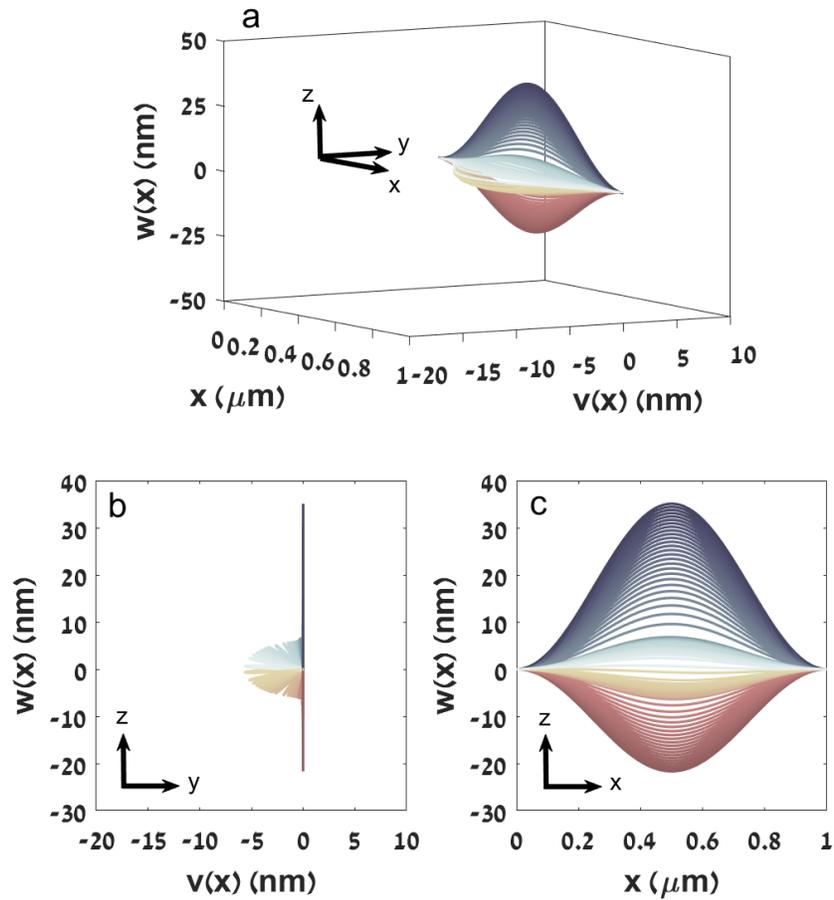

**Figure 11 Rotational motion.** *(a) 3D view, (b) side-view and (c) front-view of the rotational continuous motion from upward to downward curvatures enabled due to a small initial out-of-plane component $v_{01}$ = -1e-5.*

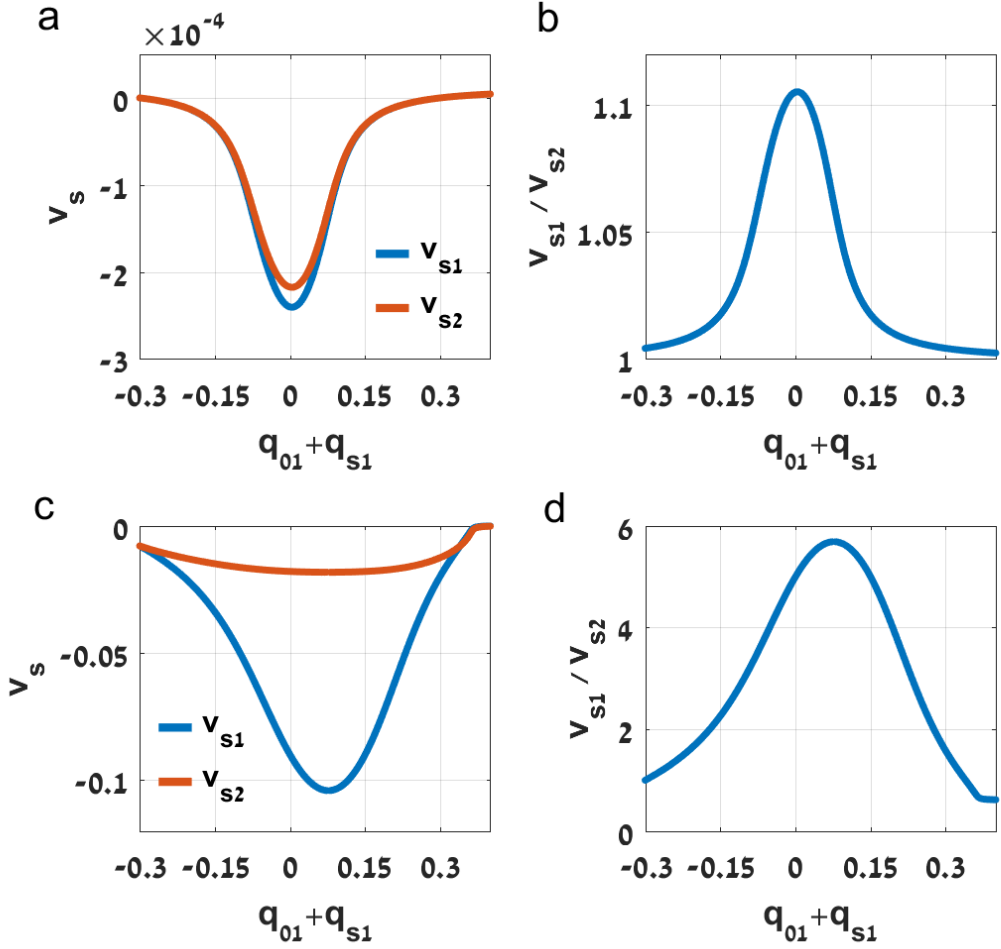

**Figure 12 Out-of-plane components evolution.** *(a) The evolution of the out-of-plane modes as a function of the in-plane symmetric deflection for initial conditions in which $v_{01,02} \ll q_{01}$. (b) The ratio between the symmetric ($v_{s1}$) and asymmetric ($v_{s2}$) out-of-plane modes from (a). (c) The evolution of the out-of-plane modes as a function of the in-plane symmetric deflection for initial conditions in which $v_{01,02} \lesssim q_{01}$. (d) The ratio between the symmetric ($v_{s1}$) and asymmetric ($v_{s2}$) out-of-plane modes from (c). While the initial components are equal ($v_{01}=v_{02}$), near the transition from upward to downward buckling, a symmetric shape becomes preferable and the ratio reaches a maximum.*

In order to achieve ST transition, we must increase the initial tension, P and the initial $q_{02}, v_{01}, v_{02}$ accordingly. Fig. 13a presents a novel three-dimensional snap-through transition. As before, out-of-plane rotational motion evolves with the electrostatic force (note the "side-view" in (b)), but at the critical point, the out-of-plane stretching is insufficient and the CNT cannot compress downward any further, exhibiting a 3D EB instability, resulting in a "jump" to a downward curvature configuration, noticeable in the "front-view" in (c).

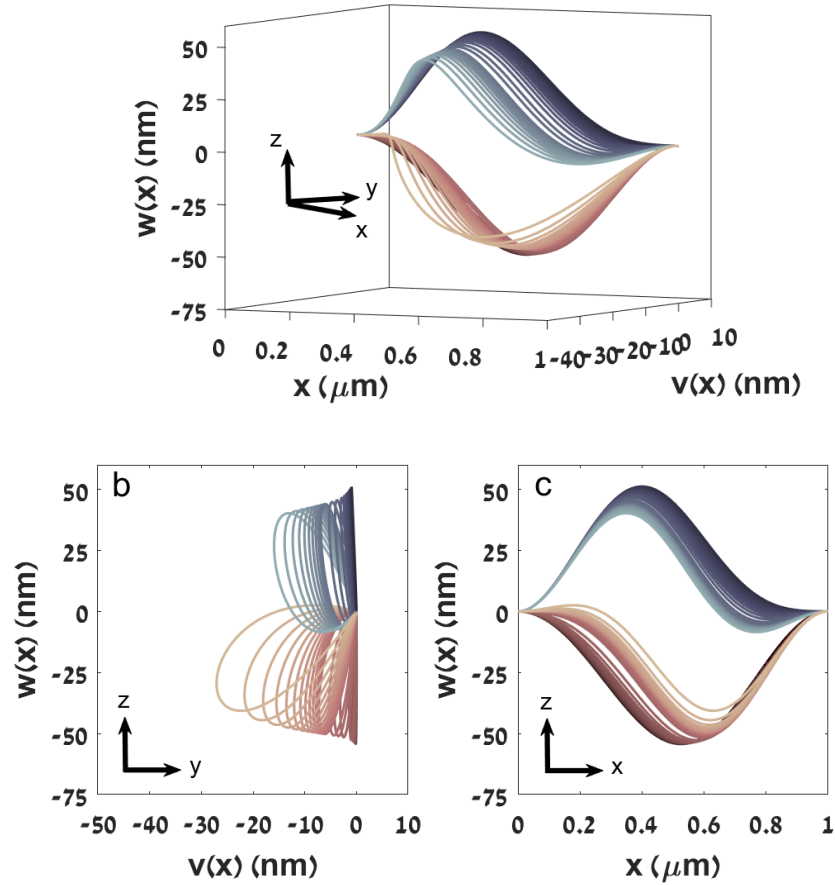

**Figure 13 Three dimensional snap-through bi-stability.** *(a) 3D view, (b) side-view and (c) front-view of a three-dimensional snap-through buckling transition.*

The evolution of the ST buckling in this case depends on five initial parameters: P, $q_{01}$, $q_{02}$, $v_{01}$, and $v_{02}$. Since we cannot visualize a five-dimensional phase space, we shall investigate a three-dimensional phase space while keeping two initial parameters fixed. First, we fix $q_{01}$ and $q_{02}$ while varying the out-of-plane initial conditions (Fig. 14a), and then we fix $v_{01}$ and $v_{02}$ and vary the in-plane initial conditions (Fig. 14b). Solving Eqs. 32 for various initial conditions (as depicted by the dots in Fig. 14 a-b), affirms that the initial configuration of the CNT determines the static response of the system. If the total tension of the CNT at zero gate voltage, (P-$\alpha Q_0$), is lower than a critical value $P_c^*$ (depicted by the lowest surface separating the blue and green dots, Fig. 14 a-b), then the CNT advances from upward to downward curvature continuously through a rotational motion (as shown in Fig. 11). When the initial total tension is higher than the critical value ($P_c^*$), the system exhibits EB ST buckling transition despite the out-of-plane movement (Fig. 13). Unlike the single-modal ST criterion (P>$P_c$=$d_{11}/b_{11}$) which is easily calculated analytically and is independent of any specific parameters of the device, an expression for $P_c^*$ is not trivial, and its value (non-dimensional) depends on the physical parameters of the device. Hence, further research needs to be conducted in order to establish the exact criteria for 3D ST buckling.

The black dots in Fig. 14a,b present a third regime (between the green and grey surfaces), in which latching bi-stability is observed[17]. In this regime, when force is applied, the CNT compresses downward ($q_{s1}$ increases), until the bifurcation point, at which the CNT "snaps" downward (marked by the top dashed arrow in Fig. 15). When the electrostatic force is removed (i.e., going back along the blue curve until the force along the z-axis equals zero), the CNT remains in downward buckling configuration. In order to achieve the release ("snap-back") transition, a negative force must be applied, meaning that a second gate electrode must be used. This configuration can be utilized for realizing a non-volatile mechanical memory element, as was previously suggested with lateral buckling configurations[20]. In order to detect the predicted latching experimentally, suitable devices with a second gate electrode must be designed. We shall note, that the initial tension, (P-$\alpha Q_0$), has a third critical value (depicted by the grey surface in Fig. 14 a-b), such that higher values result in an unstable physical solution. As in the case for the ST criteria, the latching critical value is also dependent on the physical parameters of the device, requiring further research to arrive at the general criteria for latching.

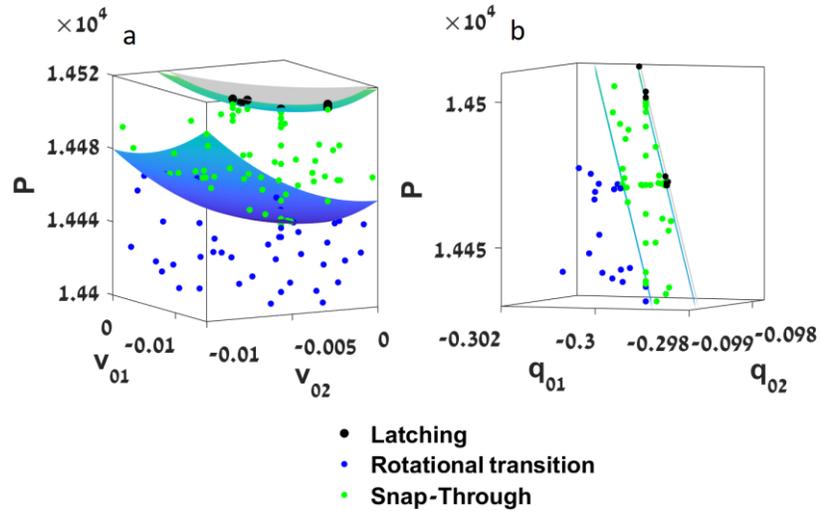

**Figure 14 Five-dimensional phase space.** *Visualization in 3D when (a) $q_{01}$, $q_{02}$ are fixed, and when (b) $v_{01}$, $v_{02}$ are fixed. The dots represent a set of parameters for which Eqs. 32 were solved, and their color represents the nature of the solution: blue for rotational continuous transition from upward to downward curvature, green for 3D ST buckling, and black for latching. The colored surfaces represent planes on which the initial non-dimensional tension ($P-\alpha Q_0$) is constant, at values separating the different regimes.*

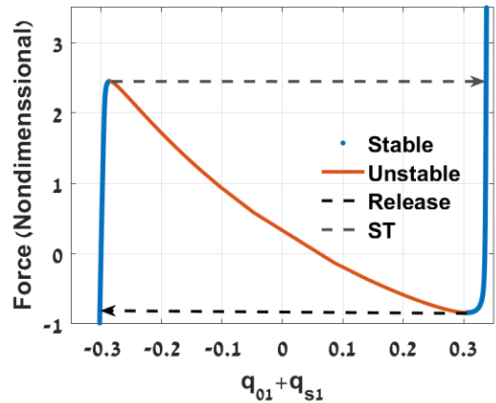

**Figure 15 Latching static response.** *Static response for initial parameters satisfying the latching criteria. Note that the release occurs at negative force, meaning that at zero gate voltage the system remains in the downward-buckled state ($q_{01}+q_{1s}>0$), and can hence serve as a non-volatile memory.*

*4.2 Eigenvalue problem analysis*

After the Galerkin decomposition, the dynamic equations (Eq. 29) are transformed to the following set of algebraic coupled equations:

$$(q_{01}+q_{s1})q_{d1}u_{111} + (q_{02}+q_{s2})q_{d2}u_{221} + q_{d1}d_{11}$$
$$+(q_{01}+q_{s1}+q_{d1})q_{d1}u_{111} + (q_{02}+q_{s2}+q_{d2})q_{d2}u_{221} +$$
$$(P-\alpha Q_{0d}) \cdot \begin{bmatrix} (q_{01}+q_{s1})q_{d1}p_{111} + (q_{02}+q_{s2})q_{d2}p_{221} - q_{d1}b_{11} \\ +(q_{01}+q_{s1}+q_{d1})q_{d1}p_{111} + (q_{02}+q_{s2}+q_{d2})q_{d2}p_{221} \end{bmatrix} -$$
$$\alpha Q_d \cdot \left[ -(q_{01}+q_{s1}+q_{d1})b_{11} + (q_{01}+q_{s1}+q_{d1})^2 p_{111} + (q_{02}+q_{s2}+q_{d2})^2 p_{221} \right] -$$
$$\omega^2 \cdot \left[ q_{d1}a_{11} + (q_{01}+q_{s1}+q_{d1})q_{d1}x_{111} + (q_{02}+q_{s2}+q_{d2})q_{d2}x_{221} \right] - e_1\kappa_d = 0$$

$$(q_{02}+q_{s2})q_{d1}u_{122} + (q_{01}+q_{s1})q_{d2}u_{212} + q_{d2}d_{22}$$
$$+(q_{01}+q_{s1}+q_{d1})q_{d2}u_{122} + (q_{02}+q_{s2}+q_{d2})q_{d1}u_{212} +$$
$$(P-\alpha Q_{0d}) \cdot \begin{bmatrix} (q_{02}+q_{s2})q_{d1}p_{122} + (q_{01}+q_{s1})q_{d2}p_{212} - q_{d2}b_{22} \\ +(q_{01}+q_{s1}+q_{d1})q_{d2}p_{122} + (q_{02}+q_{s2}+q_{d2})q_{d1}p_{212} \end{bmatrix} -$$
$$\alpha Q_d \cdot \left[ -(q_{02}+q_{s2}+q_{d2})b_{22} + (q_{01}+q_{s1}+q_{d1})(q_{02}+q_{s2}+q_{d2})(p_{122}-p_{212}) \right] -$$
$$\omega^2 \cdot \left[ q_{d2}a_{22} + (q_{01}+q_{s1}+q_{d1})q_{d2}x_{212} + (q_{02}+q_{s2}+q_{d2})q_{d1}x_{122} \right] = 0$$

$$v_{d1}d_{11} - (P-\alpha Q_{0d})v_{d1}b_{11} + \alpha Q_d(v_{01}+v_{s1}+v_{d1})b_{11} - \omega^2 a_{11}v_{d1} = 0 \qquad (33)$$

$$v_{d2}d_{22} - (P-\alpha Q_{0d})v_{d2}b_{22} + \alpha Q_d(v_{02}+v_{s2}+v_{d2})b_{22} - \omega^2 a_{22}v_{d2} = 0$$

<u>4.2.1 In-plane initial conditions</u>

Let us begin with the assumption that there is no out-of-plane initial deflection at all, meaning $v_0=0$. This results in zero out-of-plane static motion, $v_s=0$ (grey curve in Fig. 10). The in-plane and out-of-plane coupling is manifested through the tension integrals ($Q_{0d}$, $Q_d$). For $v_0=v_s=0$, only the quadratic dynamic terms preserve this coupling, represented by $Q_d^* = b_{11}(q_{d1}^2+v_{d1}^2) + b_{22}(q_{d2}^2+v_{d2}^2)$. In the linear response analysis for the resonance modes, nonlinear dynamic terms are being neglected (small vibrations), and the in-plane and out-of-plane coupling is removed. Thus, the dynamics in Eqs. 33 can be solved as two separate systems: (1) two coupled in-plane equations and (2) two uncoupled out-of-plane equations.

Let us first examine the eigenvalue problem as before:

$$\begin{pmatrix} q_{d1} & q_{d2} & v_{d1} & v_{d2} \end{pmatrix} \begin{pmatrix} H_{11} & H_{12} & 0 & 0 \\ H_{21} & H_{22} & 0 & 0 \\ 0 & 0 & H_{33} & 0 \\ 0 & 0 & 0 & H_{44} \end{pmatrix} \begin{pmatrix} q_{d1} \\ q_{d2} \\ v_{d1} \\ v_{d2} \end{pmatrix} = \omega^2 \begin{pmatrix} q_{d1} \\ q_{d2} \\ v_{d1} \\ v_{d2} \end{pmatrix} \quad (34)$$

where $H_{33} = \partial^2 H / \partial v_{d1}^2$ and $H_{44} = \partial^2 H / \partial v_{d2}^2$ (the others parameters are defined the same as in Eq. 22). The eigenvalues are extracted from the requirement that det(H)=0 and one obtains the following relation:

$$(H_{33} - \omega^2)(H_{44} - \omega^2)((H_{11} - \omega^2)(H_{22} - \omega^2) - H_{12}H_{21}) = 0 \quad (35)$$

At the critical ST transition, the Jacobian determinant must also equal zero (at the bifurcation point):

$$H_{33}H_{44}(H_{11}H_{22} - H_{12}H_{21}) = 0 \quad (36)$$

Substituting Eq. 36 into the expanded Eq. 35 eliminates the $\omega^0$ term, forcing a solution of $\omega^2=0$, which means that the zero frequency at the snap-through transition cannot be avoided. However, solving the dynamic equations reveals that the out-of-plane mode is always the lowest, and therefore the mode to reach zero frequency is actually the out-of-plane mode (Fig. 16a-b). This explains why zero resonance has never been evidenced in the experiments, since the only modes detected are the in-plane[21].

One can easily observe how all of the resonance frequencies in Fig. 16a decrease when the static force is applied due to compression. The lowest out-of-plane mode (purple) reaches zero frequency, after which there is a frequency "jump" (Fig. 16b). Applying additional force causes an increase in frequency (also known as "hardening") due to stretching of the CNT in the downward configuration. The lowest in-plane mode (yellow) and second out-of-plane mode (orange) follow the lowest out-of-plane mode, and "jump" at the same critical load from non-zero frequency (Fig. 16a,c). However, the second in-plane mode (blue) displays hardening (an increase in frequency) near the ST transition. We attribute this hardening to the fact that the transition is anti-symmetric (Fig. 13c). A symmetric vibration around the asymmetric static configuration will stretch one part of the CNT and compress the other, with an overall negligible effect on the CNT tension. However, an asymmetric vibration will stretch both parts (the upward and downward curvatures) of the tube, and therefore the anti-symmetric mode stretches the CNT, which translates into an increase of the second in-plane resonance mode.

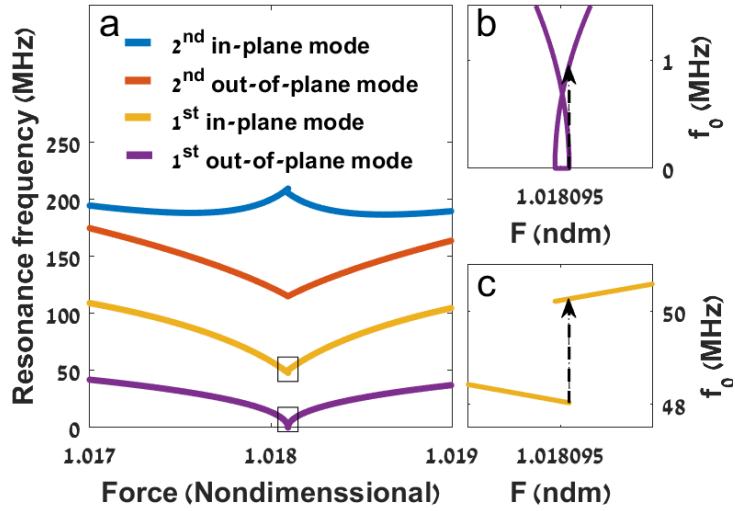

**Figure 16 Three dimensional dynamics.** *(a) Resonance frequencies gate dependence for the lowest two in-plane and lowest two out-of-plane resonance modes. Note that the out-of-plane modes are lower, and that the lowest out-of-plane mode is the one to reach zero frequency and dictate the ST "jump". (b) Zoom-in on the "jump" of the lowest out-of-plane mode when reaching zero, and (c) the consequent "jump" of the lowest in-plane mode at non-zero frequency.*

4.2.2 Realistic initial conditions

Experimentally, fabricating a device with zero out-of-plane initial deflection is extremely unlikely, effectively impossible. Also, we shall emphasize the singularity of zero initial out-of-plane conditions, which results in static motion that is only in-plane (two-dimensional). For even an extremely small out-of-plane initial component ($0<|v_0|<<<|q_0|$), static out-of-plane motion will evolve and the out-of-plane and in-plane dynamic coupling cannot be neglected, even at the linear regime.

Therefore, we must introduce small out-of-plane components to the initial conditions ($v_0 \neq 0$) as well. We solve the static equations for initial conditions which satisfy the criteria for ST buckling, as discussed in Sect. 4.1. Then, in order to find the resonance frequencies, we assume small vibrations (linear regime) and neglect dynamic nonlinear terms in Eqs. 33. We transfer the system of equations to a matrix form:

$$\underbrace{\begin{pmatrix} H_{11}-\omega^2 & H_{12} & H_{13} & H_{14} \\ H_{21} & H_{22}-\omega^2 & H_{23} & H_{24} \\ H_{31} & H_{32} & H_{33}-\omega^2 & H_{34} \\ H_{41} & H_{42} & H_{43} & H_{44}-\omega^2 \end{pmatrix}}_{M} \begin{pmatrix} q_{d1} \\ q_{d2} \\ v_{d1} \\ v_{d2} \end{pmatrix} = 0 \qquad (37)$$

and solve for det(M)=0 for every static load. Each $H_{ij}$ element represents the second derivative of the Hamiltonian with respect to the relevant coordinates. Fig. 16 presents the resulting resonance frequencies gate dependence of the suspended CNT. It can be easily seen, that just as in the case of no out-of-plane statics, the lowest resonance mode is the first out-of-plane mode. This is true both for the case of a rotational continuous transition (Fig. 17a) as well as for the case of 3D EB ST transition (Fig. 17b). The lowest out-of-plane mode (light blue) reaches zero frequency at the ST critical point and dictates the ST "jump" for all of the other modes, and thus the lowest in-plane mode (blue) follows and "jumps" at the same load from non-zero frequency.

Solving the three-dimensional equations for the resonance modes gate dependence predicts the behavior of all types of devices discussed in Ref. 21, and can be used for fitting experimental data and learn about the physical motion of the device.

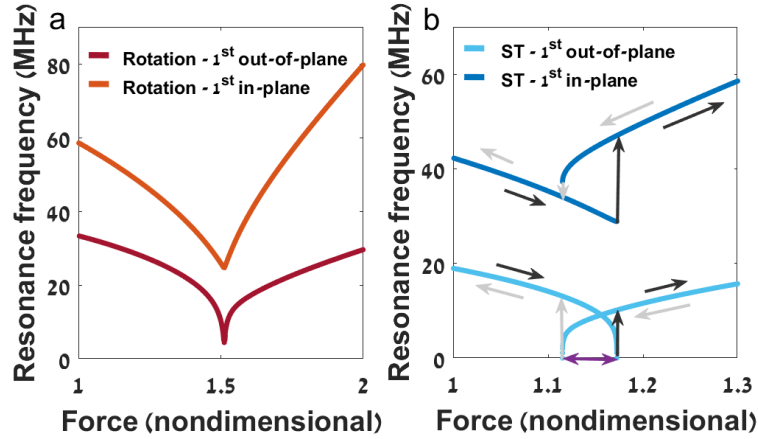

**Figure 17 Three dimensional dynamics.** *Resonance frequencies gate dependence for the full two-modals 3D analysis with realistic initial conditions for (a) a continuous rotational transition ($(P-\alpha Q_0)<P_c^*$) and (b) three-dimensional ST buckling transition ($(P-\alpha Q_0)>P_c^*$). The dark and light grey arrows present an upward or downward gate voltage sweeps, respectively, influencing the out-of-plane and in-plane modes simultaneously. Purple double-arrow represents the ST hysteresis window, i.e. the difference between static loads at which the ST (dark grey) and release (light grey) jumps occur. Out-of-plane symmetry breaking is evident, as the ST and release "jumps" (vertical arrows of the light blue curve) are not equal. Furthermore, the in-plane mode exhibits a "negative jump" (decrease in frequency) at the release (downward light grey arrow), which is occasionally evident in experimental data[21] and cannot be explained without the additional out-of-plane statics.*

## 5 Conclusion

We analyze the theoretical behavior of the recently reported CNT resonators with initial upward curvature using the framework of continuous mechanics. We model the CNT as a doubly clamped beam subjected to electrostatic force from the local gate. We use the Euler-Bernoulli beam theory to describe both the in-plane as well as the out-of-plane static and dynamic motions, analyzed using the Galerkin reduced order model. In the process of arriving at a satisfactory model, we gather several surprising insights on the mechanical system behavior. We show that the addition of a second in-plane modals cannot achieve a physical solution without adding the asymmetric modal to the initial conditions. Then we prove that a two-modals Galerkin ROM results in a satisfactory approximation, and that the electrostatic force can be rightfully approximated as inversely proportional to the displacement. Most importantly, we prove that the experimental data of snap-through buckling at finite frequency can only be explained by the addition of the out-of-plane degrees of motion. We show that the case of in-plane initial conditions is a very unique singularity, such that a realistic model must include out-of-plane initial conditions as well. We show how such initial conditions result in a hybrid three-dimensional static transition from upward to downward curvature for all types of transitions (continuous, ST or latching). Finally, we analyze the criteria for ST buckling as well as latching, which depend on the initial tension and configuration of the CNT. We believe that the model formulated in this work is of fundamental significance to the understanding of buckled CNTs, and essential for designing the next generation of buckled CNT devices for practical applications.

**Acknowledgements.** This study was supported by the ISF (Grant No. 1854/19), and the Russell Berrie Nanotechnology Institute. The work made use of the Micro Nano Fabrication Unit at the Technion. S.R. acknowledges support by the Council for Higher Education and the Russel Berrie scholarships.

**Data Availability.** Data sharing not applicable to this article as no datasets were generated or analyzed during the current study. Source codes for theoretical calculations are available from the corresponding author upon reasonable request.

***Corresponding author:** Y. E. Yaish, email: yuvaly@technion.ac.il

# Supplementary Information

## 1 Symmetric initial conditions in a two-modals static analysis

We wish to use two Galerkin modals for the in-plane analysis with a symmetric initial shape, i.e. $q_{02}=0$. Assuming $w<<g_0$, such that the electrostatic nonlinearity can be neglected, we substitute $w_0=q_{01}\varphi_1$ and $w_s=q_{s1}\varphi_1+q_{s2}\varphi_2$ into Eq. 11 and obtain:

$$\left(q_{s1}\varphi_1''''+q_{s2}\varphi_2''''\right)+\left(P-\alpha Q_0\right)\left(q_{s1}\varphi_1''+q_{s2}\varphi_2''\right)-\alpha\left(\left(q_{01}+q_{s1}\right)\varphi_1''+\left(q_{02}+q_{s2}\right)\varphi_2''\right)Q_s=\kappa_s \quad (S1)$$

where $Q_0=\int_0^1\left(q_{01}\varphi_1'\right)^2 dx=b_{11}q_{01}^2$ and

$$Q_s=\int_0^1\left[2q_{01}\varphi_1'\left(q_{s1}\varphi_1'+q_{s2}\varphi_2'\right)+\left(q_{s1}\varphi_1'+q_{s2}\varphi_2'\right)^2\right]dx=b_{11}q_{s1}^2+b_{22}q_{s2}^2+2b_{11}q_{01}q_{s1}.$$

We multiply Eq. S1 by $\varphi_1(x)$ and integrate over x from 0 to 1, and then we do the same multiplying by $\varphi_2(x)$. Eq. S1 is transformed into:

$$q_{s1}d_{11}-\left(P-\alpha b_{11}q_{01}^2\right)q_{s1}b_{11}+\alpha\left(q_{01}+q_{s1}\right)b_{11}\left(\left(2q_{01}q_{s1}+q_{s1}^2\right)b_{11}+q_{s2}^2b_{22}\right)-\kappa_s e_1=0 \quad (S2)$$

$$q_{s2}d_{22}-\left(P-\alpha b_{11}q_{01}^2\right)q_{s2}b_{22}+\alpha q_{s2}b_{11}\left(\left(2q_{01}q_{s1}+q_{s1}^2\right)b_{11}+q_{s2}^2b_{22}\right)=0 \quad (S3)$$

Assuming $q_{s2}\neq 0$, from Eq. S3 we can extract:

$$q_{s2}^2=\frac{d_{22}-\left(P-\alpha b_{11}q_{01}^2\right)+\alpha b_{22}\left(2q_{01}q_{s1}+q_{s1}^2\right)b_{11}}{-\alpha b_{22}^2} \quad (S4)$$

Substituting Eq. S4 into S2, Eq. S2 reduces to:

$$q_{s1}\left(d_{11}-\frac{b_{11}d_{22}}{b_{22}}\right)+q_{01}\left(Pb_{11}-\alpha b_{11}^2q_{01}^2-\frac{b_{11}}{b_{22}}d_{22}\right)=e_1\kappa_s \quad (S5)$$

We received a linear relation between $q_{s1}$ and $\kappa_s$, which of course cannot achieve EB ST bi-stability. Therefore, we must introduce the second modal to the initial beam configuration as well.

**2 Two modals static analysis**

Fig. S1 presents the evolution of the asymmetric mode of the CNT static shape ($q_{s2}$) as a function of the static symmetric CNT shape ($q_{01}+q_{s1}$). Negative values of the x axis represent upward buckled CNT whereas positive values represent downward buckling. Note that in the vicinity of the transition, as the symmetric mode approaches zero, the asymmetric mode is at its maximum. This correlates well to the asymmetric transition apparent in Fig. 7a of the main text, in which the CNT shape is asymmetric.

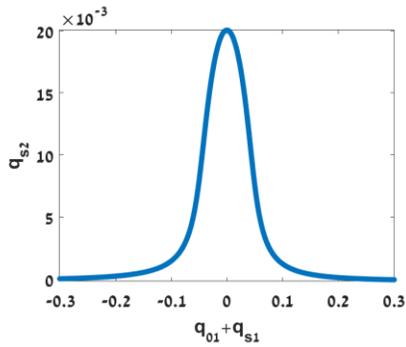

**Figure S1** *The evolution of the asymmetric mode of the CNT static shape ($q_{s2}$) as a function of the static symmetric CNT shape ($q_{01}+q_{s1}$). Negative values of the x axis represent upward buckled CNT whereas positive values represent downward buckling. Note that in the vicinity of the transition, as the symmetric mode approaches zero, the asymmetric mode is at its maximum.*